\documentclass[a4paper,twoside]{article}  \usepackage{epsfig} \usepackage{subcaption} \usepackage{calc} \usepackage{amssymb}
\usepackage{amstext}
\usepackage{amsmath}
\usepackage{amsthm}
\usepackage{multicol}
\usepackage{pslatex}
\usepackage{apalike}
\usepackage{hyperref}  
\usepackage{enumerate} 
\usepackage{enumitem}  
\usepackage[autolinebreaks,useliterate]{mcode}  
\usepackage[
activate = {true},
kerning  = true,
spacing  = true,
factor   = 1000,
stretch  = 10,
shrink   = 10,
]{microtype}
 
\renewcommand{\figurename}{Fig.}

\usepackage{SCITEPRESS}  

\begin{document}

\title{3D Augmented Reality Tangible User Interface using Commodity Hardware}

\author{\authorname{Dimitris Chamzas\sup{1}\orcidAuthor{0000-0002-4375-5281}, and Konstantinos Moustakas\sup{1}\orcidAuthor{0000-0001-7617-227X}} 
\affiliation{\sup{1}Department of Electrical and Computer Engineering, University of Patras, Rio Campus, Patras 26504, Greece}
\email{chamzas95@gmail.com, moustakas@ece.upatras.gr }
}


\abstract{During the last years, the emerging field of Augmented \& Virtual Reality (AR-VR) has seen tremendous growth. An interface that has also become very popular for the AR systems is the tangible interface or passive-haptic interface. Specifically, an interface where users can manipulate digital information with input devices that are physical objects. This work presents a low cost Augmented Reality system with a tangible interface that offers interaction between the real and the virtual world. The system estimates in real-time the 3D position of a small colored ball (input device), it maps it to the 3D virtual world and then uses it to control the AR application that runs in a mobile device. Using the 3D position of our ``input'' device, it allows us to implement more complicated interactivity compared to a 2D input device. Finally, we present a simple, fast and robust algorithm that can estimate the corners of a convex quadrangle. The proposed algorithm is suitable for the fast registration of markers and significantly improves performance compared to the state of the art.}

\onecolumn \maketitle \normalsize \setcounter{footnote}{0} \vfill

\section{\uppercase{Introduction}}
\label{sec:introduction}

\noindent Augmented \& Virtual Reality (AR-VR) systems and applications have seen massive development and have been studied extensively over the last few decades \cite{azuma1997survey,billinghurst2015survey,moustakas2015}.
Virtual reality (VR) is an artificial 3D environment generated with software. Users are immersed in this 3D world, and they tend to accept it as a real environment. On the other hand, Augmented Reality (AR) is a technology that blends digital content into our real world. Thus, AR combines real and virtual imagery, is interactive in real-time, and registers the virtual imagery with the real world. AR \& VR systems require specialized hardware and most of the time they are quite expensive.

In contrast with Virtual Reality, where the user is completely immersed in a virtual environment, AR allows the user to interact with the AR digital world and manipulate the virtual content via special input devices. Three-dimensional visualization would be ideally accompanied by 3D interaction, therefore 3D input devices are highly desirable \cite{reitmayr2005iorb}. To reduce the complexity, the input device we choose, is a simple, colored ball at the end of a stick with three degrees of freedom (DOF). We will refer to it as the \textsc{ar-pointer}. 
  
Building an AR system we have to decide on how to implement its three basic functions. \textit{Display}, where we have to combine images from the real and virtual world, \textit{Tracking}, where we have to find the position of the user's viewpoint in the real world and register its view in the 3D virtual world and a \textit{User Interface} (Input-Interaction), where a computer responds in real-time to the user input and generates interactive graphics in the digital world. 
With the advances in mobile device technology, handheld computing devices are becoming powerful enough to support the functionalities of an AR System.  Google's ARcore platform is such an example.  Considering that we want to build a low-cost AR system with a 3D tangible input interface, we choose a mobile phone, running Unity and Vuforia, to implement the AR Video-based \textit{Display} and the \textit{Traking} modules while the tangible \textit{User Interface} is implemented in a "custom-made" device running Open Source software. 

The contributions of this paper are threefold.
First, we describe the development of a DIY (Do It Yourself) low-cost AR-VR working prototype system with a 3D tangible user interface that can be used as a test-bed to examine a variety of problems related to 3D interaction in VR or AR environments. Second, the usage of the real 3D position of the tangible input device obtained via an adaptive color and distance camera registration algorithm, offers a powerful and flexible environment for interactivity with the digital world of AR-VR. Third, we present cMinMax, a new algorithm that can estimate the corners of a convex polygon. This algorithm is suitable for the fast registration of markers in augmented reality systems and in applications where real-time feature detector is necessary. cMinMax is faster, approximately by a factor of 10, and more robust compared to the widely used Harris Corner Detection algorithm.
\section {\uppercase{Related Work}} \label{sec:Related}
\noindent During the last two decades AR research and development have seen rapid growth and as more advanced hardware and software becomes available, many AR systems with quite different interfaces are moving out of the laboratories to consumer products. 
Concerning the interactivity of an AR system, it appears that users prefer for 3D object manipulation to use the so-called Tangible User Interface (TUI) \cite{billinghurst2008tangible,ishii2008tangible},
Thus for the Interactivity interface, we follow the Tangible Augmented Reality approach, a concept initially proposed by T. Ishii \cite{ishii1997tangible}. 
This approach offers a very intuitive way to interact with the digital content and it is very powerful since physical objects have familiar properties and physical constraints, therefore they are easier to use as input devices \cite{ishii2008tangible,shaer2010tangible}.
Besançon et al. compared the mouse-keyboard, tactile, and tangible input for AR systems with 3D manipulation \cite{besanccon2017mouse}. They found that the three input modalities achieve the same accuracy, however, tangible input devices are more preferable and faster. 
To use physical objects as input devices for interaction requires accurate tracking of the objects, and for this purpose, many Tangible AR applications use computer vision-based tracking software. 

An AR system with 3D tangible interactivity and optical tracking is described in \cite{martens2004experiencing}, where 3D input devices tagged with infrared-reflecting markers are optically tracked by well-calibrated infrared stereo cameras. Following this approach, we attempted to build an AR system with a 3D tangible input interface using a multicamera smartphone. Unfortunately, it did not succeed because neither Android or iOS SDKs were offering adequate support for multiple cameras nor the smartphone we used was allowing full access to their multiple cameras images to estimate the distance of the input device using binocular stereo vision. In addition, the fact that the cameras were too close and not identical it was one more problem.

With the recent technical advances and commercialization of depth cameras (e.g. Microsoft Kinect) more accurate tracking of moving physical objects became available for VR-AR applications. Such an approach is described in \cite{hernandez2012detecting} where the 3D position of a moving object is estimated utilizing the images of an RGB camera and a depth sensor. Taking into consideration that depth cameras start to appear in mobile devices, we decided to follow a similar approach and instead of using stereo vision to estimate the distance of the 3D input device we use one RGB and one depth camera.

A different approach is used in \cite{teng2017augmented}, where a user with a mobile device and two AR "markers" can perform a 3D modeling task using a tangible interface. Markers are realized as image targets. The first image target is applied to create the virtual modeling environment and the second image target is used to create a virtual pen. Using Vufuria's platform they estimate its position in the 3D world and interact accordingly. However, this input device, a stick with an image target attached to its end, is difficult to use, it is not accurate and it is not a real 3D input device since the system knows its 3D position in the virtual world but not in the real one. 
\section{\uppercase{System Architecture}} \label{sec:System}
 
 \noindent
Our system architecture implements the three modules of an AR-system, \textit{Tracking}, \textit{Display} and \textit{User Interface},
\begin{figure}[ht]
\centering
\includegraphics[width=1.0\columnwidth]{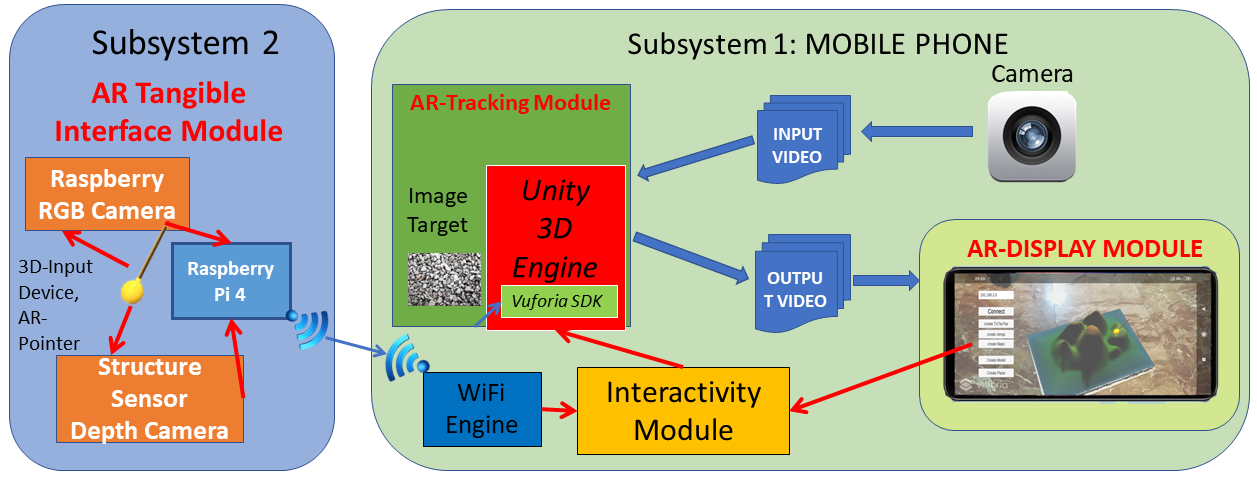}
\caption{System Architecture.}
\label{fig:archit}
\end{figure}
in two separate subsystems that communicate via WiFi (\autoref{fig:archit}).

The \textbf{\textit{first subsystem}} implements \textit{Tracking} and \textit{Display}, and it contains an Android mobile phone (Xiaomi Red-mi Note 6 pro)
\begin{figure}[ht]
\centering
\includegraphics[width=1.0\columnwidth]{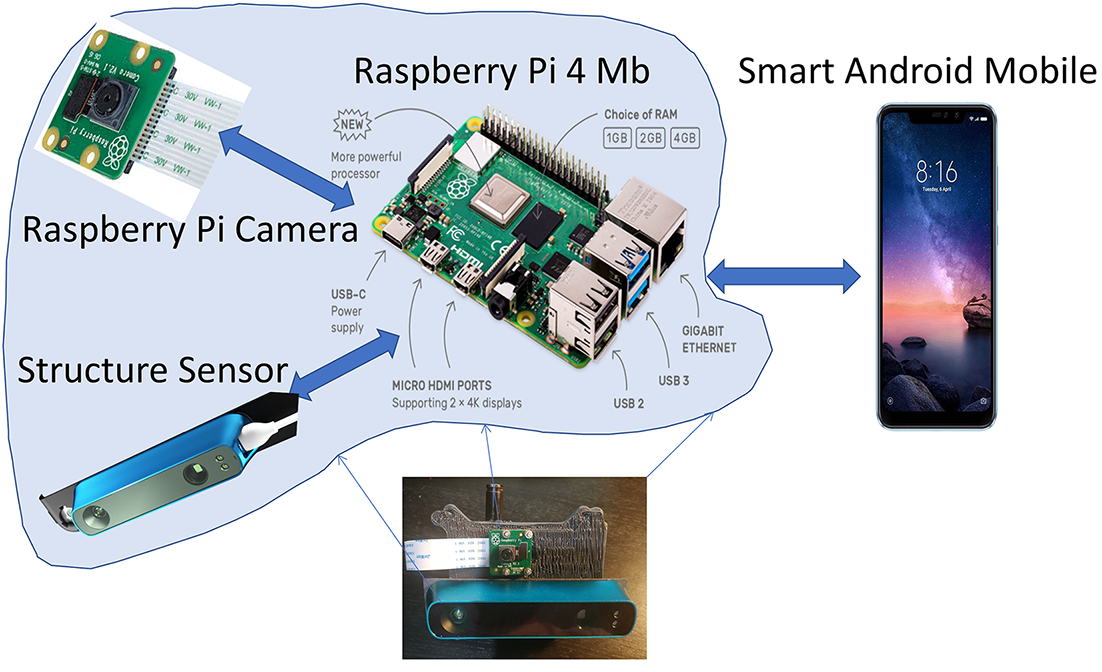}
\caption{Hardware.}
\label{fig:hard1}
\end{figure}
(\autoref{fig:hard1}) running a 3D Unity Engine with Vuforia  where  by utilizing an "image target" the front camera projects the augmented environment on our screen. 

The \textbf{\textit{second subsystem}} implements the 3D tangible AR \textit{User's interface} (TUI) and it consists of a Raspberry Pi 4 with an RGB Raspberry Camera and a depth camera (Structure Sensor) housed in a homemade 3D-printed case. The Structure Sensor projects an infrared pattern which is reflected from the different objects in the scene. Its IR camera captures these reflections and computes the distance to every object in the scene while at the same time the Raspberry camera captures the RGB image.
All the processing power in the second subsystem is done in the Raspberry Pi 4. It uses python as well as the OpenCV library for image processing, Matlab was also used as a tool for testing main algorithms before their final implementation.

\section{\uppercase{Implementation}} \label{sec:Implement}
\begin{figure}[h]
\includegraphics[width=1.0\columnwidth]{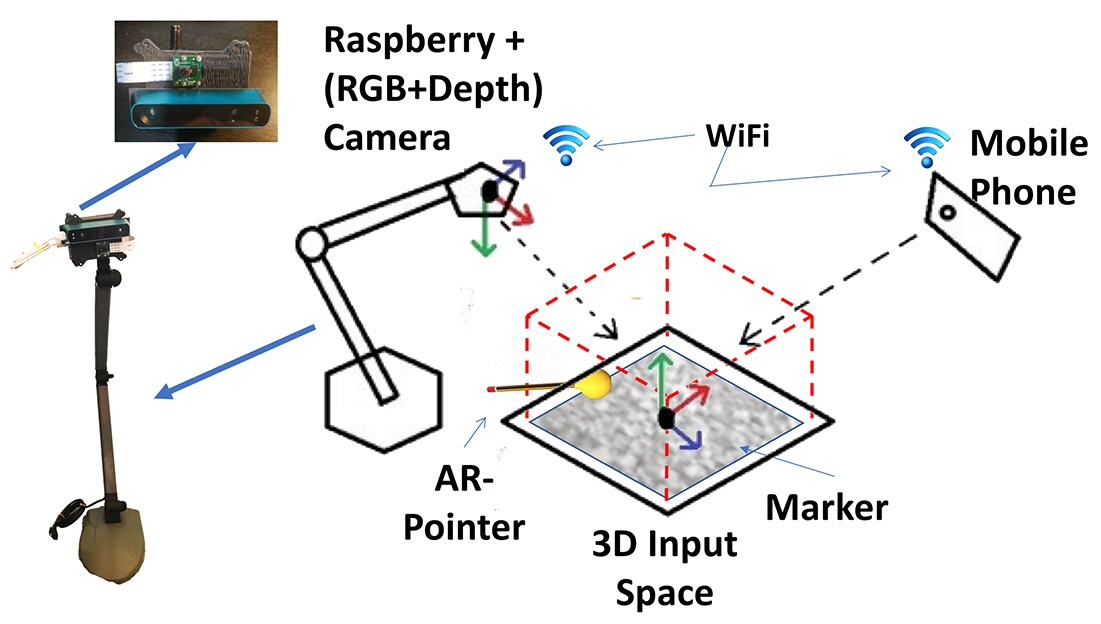}
\caption{The AR system.}
\label{fig:setup}
\end{figure}

\begin{figure}[ht]
\centering
\includegraphics[width=.99\columnwidth]{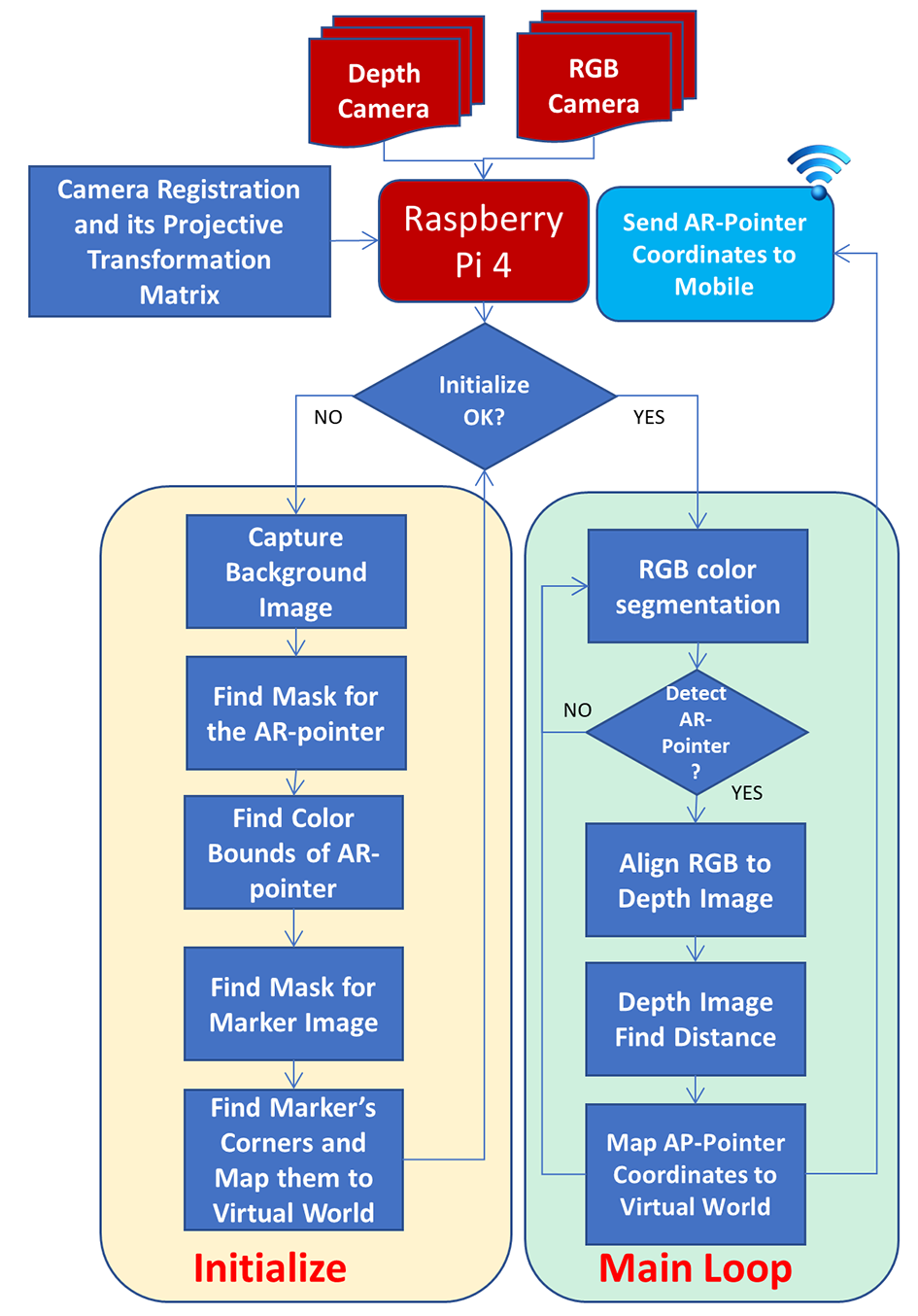}
\caption{Flow Chart of Processes Running in Raspberry.}
\label{fig:flow1}
\end{figure}

\noindent An illustrative example of the setup of our system in the real world is shown in \autoref{fig:setup}.
As it was described in \autoref{sec:System} our system is composed of two different subsystems.
The first subsystem, the mobile phone, is responsible for the visualization of the Augmented Reality environment. The virtual objects are overlaid on a predefined target image printed on an A4 paper. Thus the mobile phone is responsible for graphics rendering, tracking, marker calibration, and registration as well as for merging virtual images with views of the real world. 
The second subsystem is attached to a "desk lamb arm", and faces the target image. This system is responsible to align the images from the two cameras, locate the 3D coordinates of a predefined physical object(yellow ball), namely the \textsc{ar-pointer}, transform its XYZ coordinates to  Unity coordinates and send them to mobile via WiFi. The physical pointer, which has a unique color, is localized via an adaptive color and distance camera registration algorithm.
The physical \textsc{ar-pointer} has its virtual counterpart in the  AR world, a virtual red ball, which represents the real 3D input device. Thus, by moving the real \textsc{ar-pointer} in the real 3D world, we move the virtual \textsc{ar-pointer} in the virtual world, interacting with other virtual objects of the application. This is a tangible interface, which gives to the user the perception of a real object interacting with the virtual world. The subsystems use the marker as the common fixed frame and communicate through wi-fi.  \autoref{fig:archit} shows the building blocks of our subsystems and \autoref{fig:flow1} displays the flowchart of the processes running in the second subsystem.

\subsection{Camera Registration} \label{subsec:Camera}

\noindent Since the two different cameras (RGB, Depth) are connected to the Raspberry and the position of each other is different in space, the images taken by the two cameras are slightly misaligned. 
To correct this, we find a homographic transformation that compensates differences in the geometric location of the two cameras. Using a plug-in of matlab called registration-Estimator we select SIFT algorithm to find the matching points and to return affine transformation.

\subsection{Initialization}  \label{subsec:Init}

\noindent At initialization, masks that filter the background, the color bounds of the physical \textsc{ar-pointer} and the real to virtual world coordinates mappings are calculated.

\subsubsection{Find Mask}  \label{subsub:FindMask}
\noindent During initialization, the image target, and the \textsc{ar-pointer}) need to be separated from their background. To this end, two binary masks are created to filter out the background with the following method: 
\begin{enumerate}
    \item Capture background image.
    \item Place object (Marker or \textsc{ar-pointer}) and capture the second image.
    \item Subtract images in absolute value.
    \item Apply adaptive threshold (Otsu) \& blur filter.
    \item Edge detection ( Canny algorithm) \& fill contours. 
    \item Create a binary image (mask) by selecting the contour with the largest area.
\end{enumerate}

\begin{figure}[h]
\includegraphics[width=.32\columnwidth]{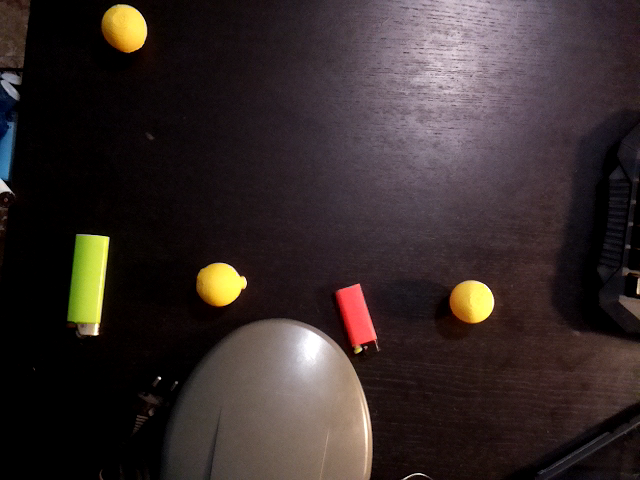}
\hfill
\includegraphics[width=.32\columnwidth]{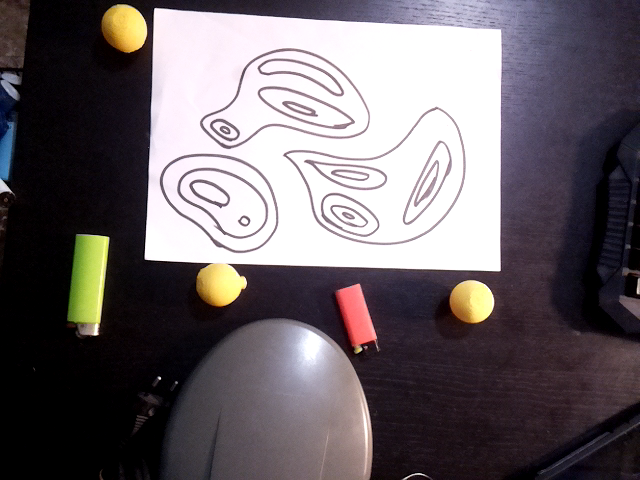}\hfill
\includegraphics[width=.32\columnwidth]{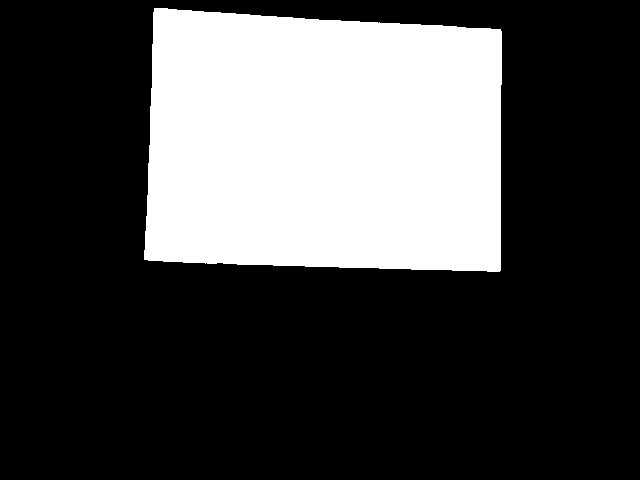}
\caption{The background without and with the object and the mask. }
\label{fig:createMaskPointer}

\end{figure}

In \autoref{fig:createMaskPointer} we see the inputs to create the mask (steps 1 and 2) and the obtained mask (step 6)

\subsubsection{Find Color Bounds}
\label{subsec:ColorBounds}

\noindent Variations in the room illumination can make the same object appear with different RGB values. To address this, the following method was developed that detects the color bounds of the \textsc{ar-pointer} under the light conditions of the room. The decided to use the HSV representation since the colors are separable in the HUE axis whereas in RGB  all the three axes are needed.    

\begin{enumerate}
    \item Find mask of \textsc{ar-pointer} (see \ref{subsub:FindMask}).
    \item Isolate pointer from image.
    \item Convert RGB image to HSV.
    \item Calculate histogram \& find max value of Hue HSV.
    \item Create new bounds $\pm 15$ at HSV.
\end{enumerate}

\begin{figure}[ht]
\centering
\includegraphics[width=.85\columnwidth]{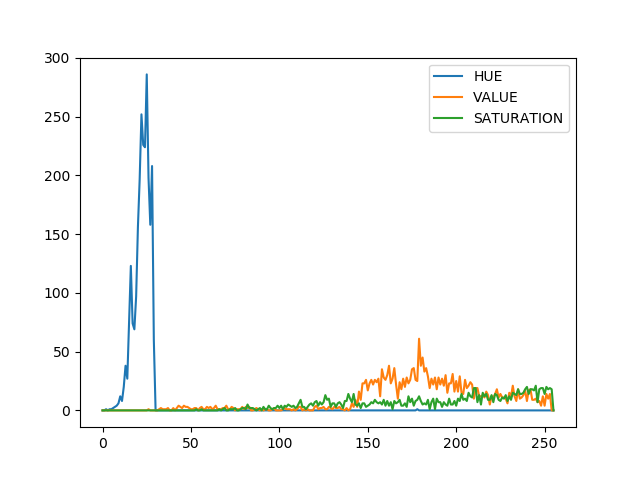}
\caption{HSV Histogram.}
\label{fig:histogram2}
\end{figure}

\begin{figure}[ht]
\centering
\includegraphics[width=.40\columnwidth]{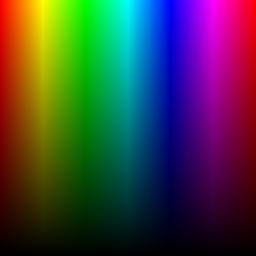}
\hfill
\includegraphics[width=.40\columnwidth]{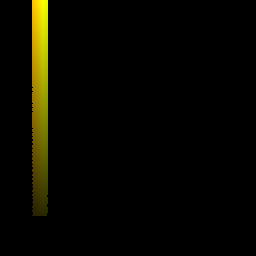}
\caption{Derived bounds isolate the yellow spectrum.}
\label{fig:colorSpectrurm2}
\end{figure}

In \autoref{fig:histogram2} the histogram of the example in \autoref{fig:WithRegister} is shown. It can be derived (step 4) that the \textsc{ar-pointer} is near Hue=20 \footnote{8bit pixel value, to obtain real HUE values multiply  by 2, since in OpenCV  max HUE is 180}.
By applying the derived bounds $(5 \leq \text{HUE} \leq 35)$ on the  RGB color spectrum only the  yellow spectrum is kept  (\autoref{fig:colorSpectrurm2}).
Identifying the color bounds makes our system robust to light variations and enables the use of multiple differently colored \textsc{ar-pointer} objects.


\subsubsection{AR Registration}  \label{subsec:ARreg}
 To allow the interaction with the digital world via the motion of the \textsc{ar-pointer} we need to map its real-world 3D coordinates $(x_r, y_r, z_r)$ to the digital world coordinates $(x_v, y_v, z_v)$. To calculate this mapping the common object of reference is the Image Target as shown in \autoref{fig:setup}.

Since the image frames are almost vertically aligned  we can approximate the relation between the z coordinates (distance) with a scalar factor $\rho _ z \approx \rho _z (x,y)$ which is proportional to the size of the image (see also \ref{subsub:MapRealVirtual}). This can be derived to obtain  $z_v= \rho_z z_r $. To map the $(x_r ,y_r )$ coordinates we need to find a  projective transformation matrix ($T_{RV}$) to account mainly for translation and rotation offset. To calculate $T_{RV}$, we need at least four points. To this end, we used the four corners of the image target mask, and map them to the Unity four corners of the marker  (\autoref{fig:RealVirtual}). Unity has constant coordinates where each corner of the marker is located at $(\pm 0.5, \pm 0.75)$, So first we will find the corners of the marker running the appropriate software in Raspberry and then we will find the transformation matrix that moves those points to the Unity virtual space.

\begin{figure}[ht]
\includegraphics[width=.45\columnwidth]{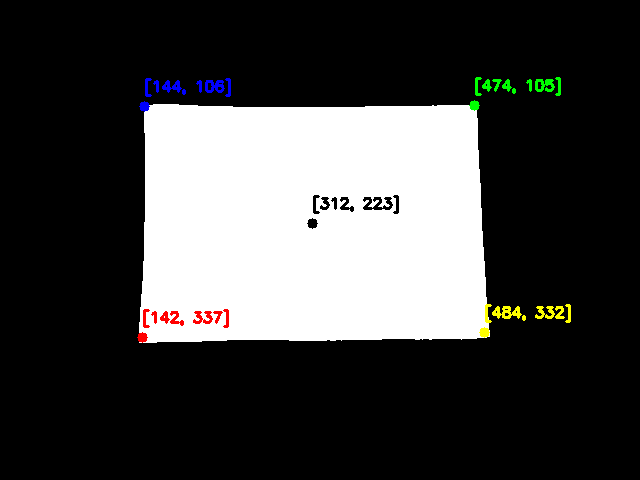}
\hfill
\includegraphics[width=.45\columnwidth]{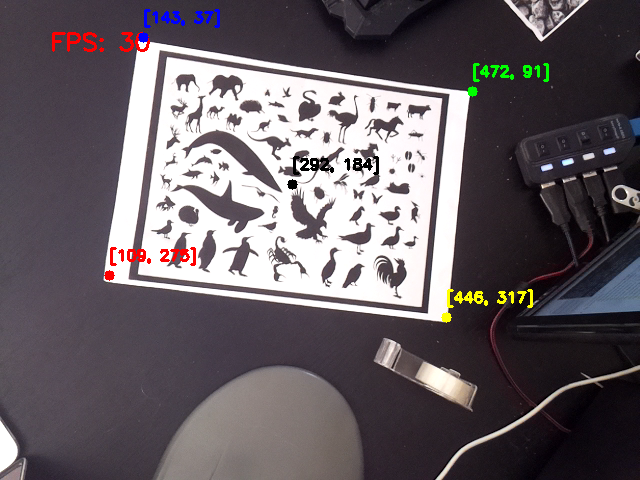}
\caption{Corner Detection.}
\label{fig:matchWorlds}
\end{figure}

To find the corners of the marker we first create the mask of the marker as described in \ref{subsub:FindMask} and then find its corners.  Initially, we used the Harris Corner Detection Algorithm from OpenCV \cite{opencv03}, but later on, we developed another simpler and faster algorithm, the cMinMax (see \autoref{subsec:cMinMax}). After we found the four corners (see \autoref{fig:matchWorlds}) we add a fifth point, the center of gravity of the marker for better results. We calculate the center of gravity as the average of X \& Y coordinates for all the pixels in the mask.

Now,  we use the two sets of points (Real World points, Unity points) and with the help of OpenCV, we get the projective transformation matrix. 
This process needs to be done only once at the start of the program and not in the main loop gaining in computational power. The result of matching the real world with the virtual one is that we can now project the virtual \textsc{ar-pointer} at the same position where the real \textsc{ar-pointer} is on the smartphone screen, making those 2 objects (real-yellow,virtual-red) to coincide.

\subsection{cMinMax: A Fast Algorithm to Detect the Corners in a Quadrangle}
\label{subsec:cMinMax}
\noindent
A common problem in image registration (see section \ref{subsec:ARreg}) is to find the corners of an image. One of the most popular algorithms to address this problem is the Harris Corner Detection \cite{harris1,opencv03}. However, most of the time the image is the photo of a parallelogram, which is a convex quadrangle. To address this specific problem we have developed a specific algorithm, referred to as \textit{cMinMax}, to detect the four corners in a fast and reliable way. The algorithm utilizes the fact that if we find the x-coordinates of the pixels that belong to the mask, then their maximum, $x_{max}$, is a corner's coordinate. Similarly for $x_{min}$, $y_{min}$ and $y_{max}$. The proposed algorithm is approximately 10 times faster and more robust than the Harris Corner Detection Algorithm, but its applicability is limited only to convex polygons. 

The basic steps of the algorithm are:
\begin{enumerate}
    \item \textbf{Preprocessing:} Generate a bi-level version of the image with the mask.
    \item  Project the image on the vertical and horizontal axis and find the $( x_{min} , x_{max} , y_{min} , y_{max} )$. These are coordinates of four corners of the convex polygon.
    \item If $N$ is the expected maximum number of angles, then for $k=1,..,int(N/2)-1$, rotate the image by $\Delta \theta = k *  pi /N $ and repeat the previous step. Identify the four additional corners, rotate the image backward by $- \Delta \theta$ and find their position in the original image.
   \item In the end, we have found $ 2N $ points which is greater than the number of expected polygon corners. Hence, there are more than one pixels around each corner. The centroids of these bunches are the estimated corners of the convex polygon.
    \item If the number of detected corners is less than N, repeat the previous three steps by rotating the image with $\Delta \theta = (k *  pi /N)- pi/2N $ 
\end{enumerate}

 \begin{figure}[h!]
  \centering
   \includegraphics[width=.90\columnwidth]{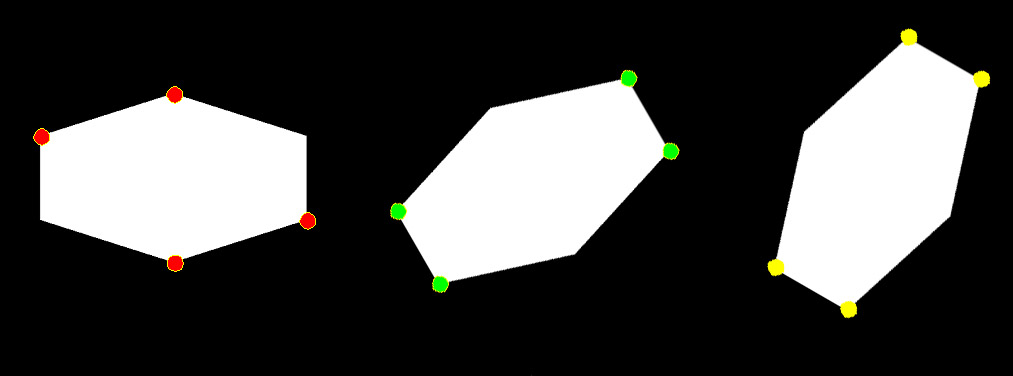}
  \caption{Detected corners in a hexagon for M=3.}
  \label{fig:hexagon_rotate}
 \end{figure}
 
 In \autoref{fig:hexagon_rotate} we apply the algorithm in a hexagon and we find all the corners with three rotations.


\noindent  
\subsection{\textsc{ar-pointer} detection (Main Loop)} \label{subsec:ARpointer}
In the main loop the 3D position of the physical \textsc{ar-pointer} is continuously estimated and transmitted to the mobile phone.

\subsubsection{RGB Color Segmentation}  \label{subsub:RGBSegm}
\noindent The X \& Y coordinates will be acquired from the RGB image. The \textsc{ar-pointer} that we used is a "3D-printed" yellow ball. Our segmentation is based on color, thus we can use any real object with the limitation to have a different color from the background. Since we have our HSV color bounds (see section \ref{subsec:ColorBounds}) the detection of the object used as the \textsc{ar-pointer} is straightforward.

\begin{enumerate}
    \item Convert RGB input image to type HSV.
    \item Keep pixel values only within preset HSV color bounds (section \ref{subsec:ColorBounds}).
    \item Edge detection ( Canny algorithm). 
    \item Find and save the outlined rectangle of contour with maximum area.
    
\end{enumerate}

\begin{figure}[h]
\centering
\includegraphics[width=.45\columnwidth]{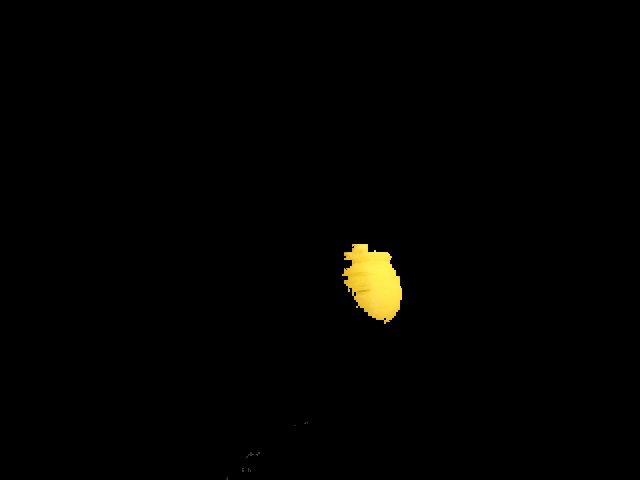}
\hfill
\includegraphics[width=.45\columnwidth]{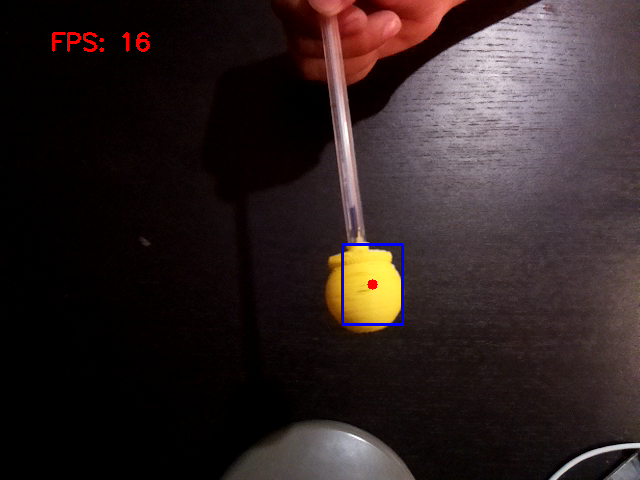}

\caption{Color detection.}
\label{fig:isolation2}
\end{figure}

Filtering the RGB image with the color bounds results to the first image of \autoref{fig:isolation2}. Then, we create a rectangle that contains the \textsc{ar-pointer} and use its center as the coordinates of it (see the second image of \autoref{fig:isolation2}).

\subsubsection{Depth Estimation} \label{subsub:DepthFind}

\begin{figure}[h]
\includegraphics[width=.45\columnwidth]{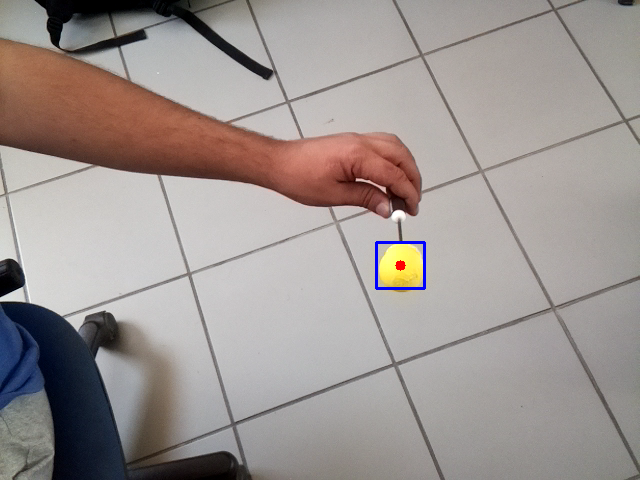}
\hfill
\includegraphics[width=.45\columnwidth]{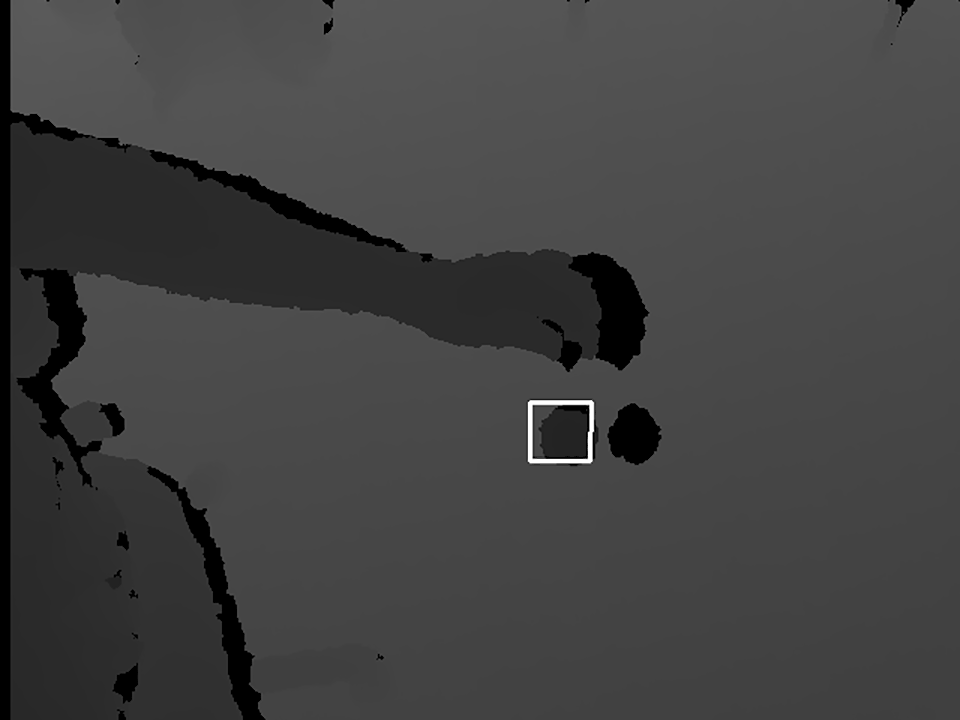}
\caption{\textsc{ar-pointer} 3D detection}
\label{fig:WithRegister}
\end{figure}

\begin{figure}[h]
\centering
\includegraphics[width=.75\columnwidth]{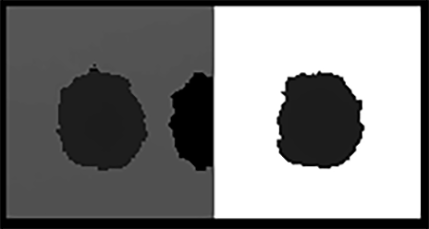}
\caption{Pre-process of the depth image.}
\label{image:trackerDepth2}
\end{figure}

The 3D coordinates of the \textsc{ar-pointer} will be acquired from the depth image. Knowing where the \textsc{ar-pointer} is located in the RGB image from the Color Segmentation, and since the images are aligned, we crop a small rectangle from the depth image that contains the area of the \textsc{ar-pointer} (\autoref{fig:WithRegister}).
This rectangle contains all the depth information we need and since it is a small part of the image it reduces also the computational cost. In this rectangle there are 3 different depth information (see \autoref{image:trackerDepth2}):
\begin{enumerate}
    \item Depth information of the \textsc{ar-pointer} (pixel values: 1000-8000).
    \item Depth information of background (pixel values: 7000-8000).
    \item Depth information for the area which is created by the \textsc{ar-pointer}  that blocks the IR emission creating a shadow of the object (pixel value: 0 ).
\end{enumerate}

\noindent Given the fact that the background always corresponds to the maximum value, we do the following on.

\begin{enumerate}
    \item We calculate the average of non-zero elements of rectangle image, (\autoref{image:trackerDepth2} first image).
    \item Set to zero all pixels with values $> 10 \%$ of average, (\autoref{image:trackerDepth2} second image).
    \item Recalculate average of non-zero elements and use it as the final depth value for the \textsc{ar-pointer}.
\end{enumerate}

With this approach, we get stable values for small changes of \textsc{ar-pointer}, fast results and precision below 1 cm.

\subsubsection{Map \textsc{ar-pointer} Coordinates to Virtual Word} \label{subsub:MapRealVirtual}

\begin{figure}[h]
\includegraphics[width=.95\columnwidth]{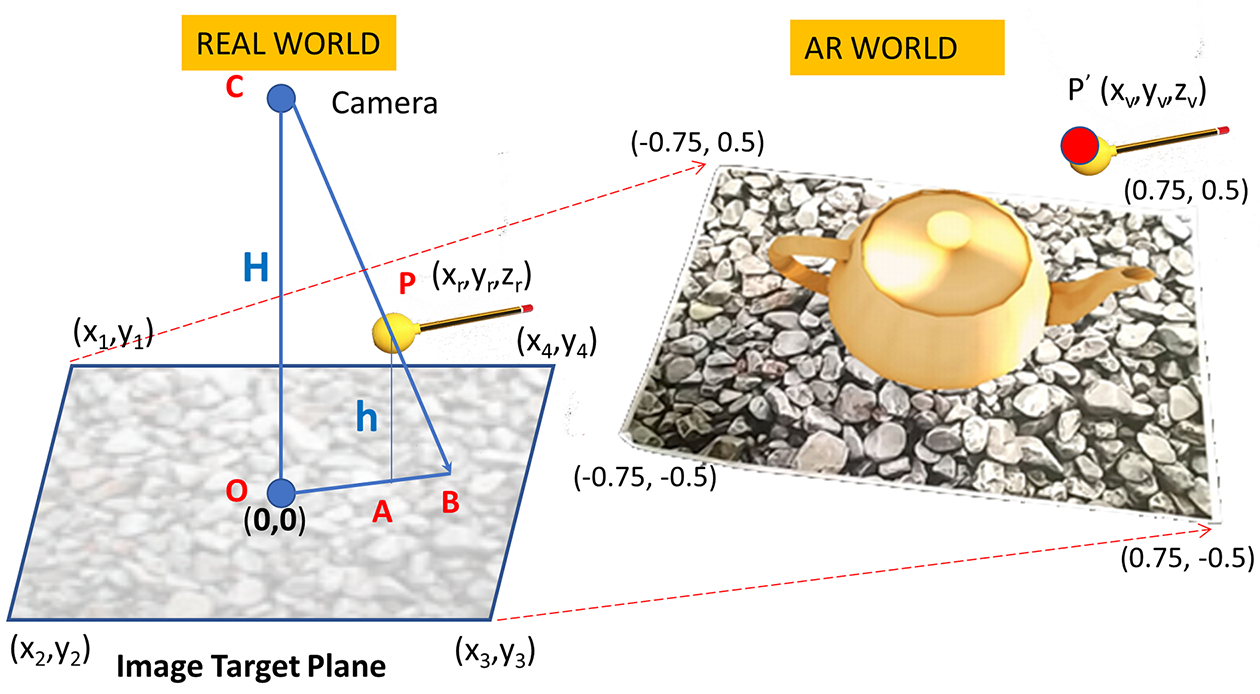}
\caption{Depth Correction.}
\label{fig:RealVirtual}
\end{figure}

\noindent At this point we know the distance of the \textsc{ar-pointer} from the Image Target plane (see \autoref{subsub:DepthFind}), as well its position on the RGB image (see \autoref{subsub:RGBSegm}). Since the \textsc{ar-pointer} is not in the Image Target plane, the position of the \textsc{ar-pointer} on the RGB image is the B point and not the A (see \autoref{fig:RealVirtual} ) as it should be. We know 
$h$ and $H$, therefore the correction vector $\overrightarrow{(AB)}$ is given from the relation $ \overrightarrow{(AB)}=  \overrightarrow{(OA)}*h/H$. 

Therefore the coordinates of the \textsc{ar-pointer} in the real word are 
\begin{align*}
x_{rA}&=x_{rB} *\frac{ (OB)-(AB)}{(OB)} \\ 
y_{rA}&=y_{rB} * \frac{(OB)-(AB)}{(OB)}, \hspace{0.5cm} 
z_{r}=h
\end{align*}

\noindent and the coordinates in the virtual world are 
\begin{center}
$
\begin{bmatrix}
x_{vA}\\ 
y_{vA}\\ 
1
\end{bmatrix}
=  T_{RV}
\begin{bmatrix}
x_{rA}\\ 
y_{rA}\\ 
1
\end{bmatrix}
$
, $z_{v} = \rho_z * z_{r}$
\end{center}

\noindent where $T_{RV}$ and $\rho_{z}$ were define in \autoref{subsec:ARreg}.
\subsection{AR Engine} \label{subsec:ArEng}
\noindent The final AR engine is a marker-based AR-system with AR video display.
It runs exclusively in the mobile phone, \autoref{fig:archit}, and is based on the 3D Unity platform.   It executes the following steps.
\begin{enumerate}
    \item Capture images with the mobile's built in camera.
    \item Detects the image target(marker) in the real world.
    \item Displays the virtual environment on top of the image target and the virtual \textsc{ar-pointer}  in the mobile screen. 
\end{enumerate}


\section{Applications and Experiments} \label{sec:App}
\noindent Using the tangible user interface three AR application were developed. The 3D position of the \textsc{ar-pointer} is used to interact with the virtual world where it appears as a red ball. 

\subsection{ The Tic-Tac-Toe game application}
\noindent A Tic-Tac-Toe  game was implemented on top of our system to demonstrate its interactivity. In \autoref{fig:TicTac}, where a screenshot of the application is shown, the option of selecting and deselecting an object (such as X or O) into the Virtual world is highlighted. It can be seen that our system offers the simple but important commands of that every AR application does require.

\subsection{ The Jenga game application}
\begin{figure}[h]
\centering
 \begin{minipage}{0.53\columnwidth}
\includegraphics[width=\linewidth]{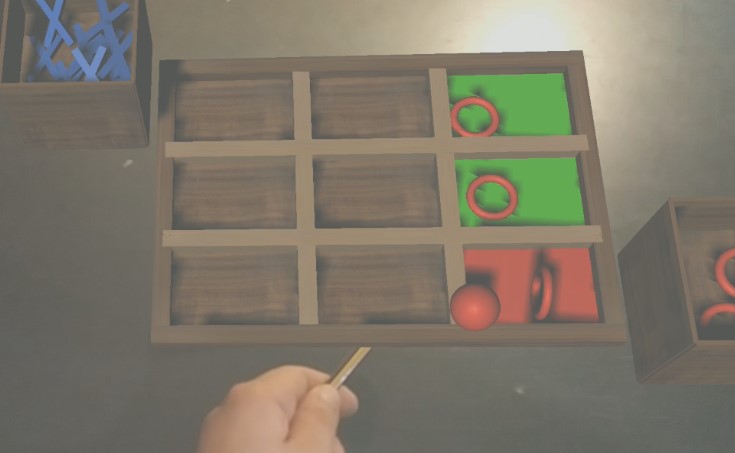}
\caption{Tic-Tac-Toe Game.}
\label{fig:TicTac}
\end{minipage}
\begin{minipage}{0.43\columnwidth}
\includegraphics[width=\linewidth]{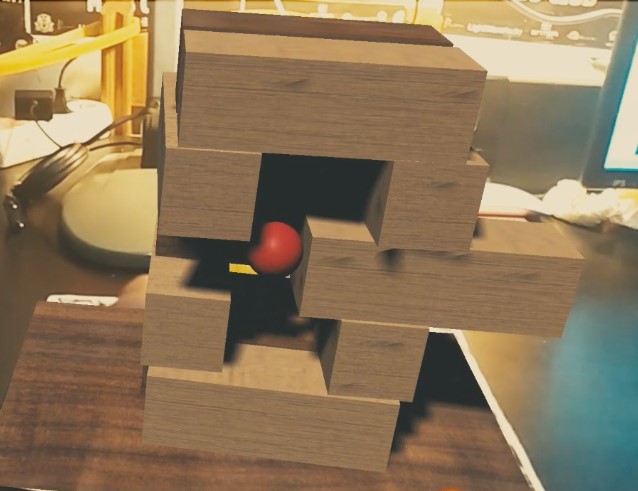}
\caption{Jenga Game.}
\label{fig:Jenga}
 \end{minipage}
\end{figure}

\noindent Additionally, a Jenga game was implemented to demonstrate the precision and stability of our system. This application can be seen in  (\autoref{fig:Jenga}. Since this game demands such features in real life, its implementation in our system, showcases the system's practicality and functionality.

\subsection{ The 3D Contour Map application}
\begin{figure}[h]
\centering
\includegraphics[width=.95\columnwidth]{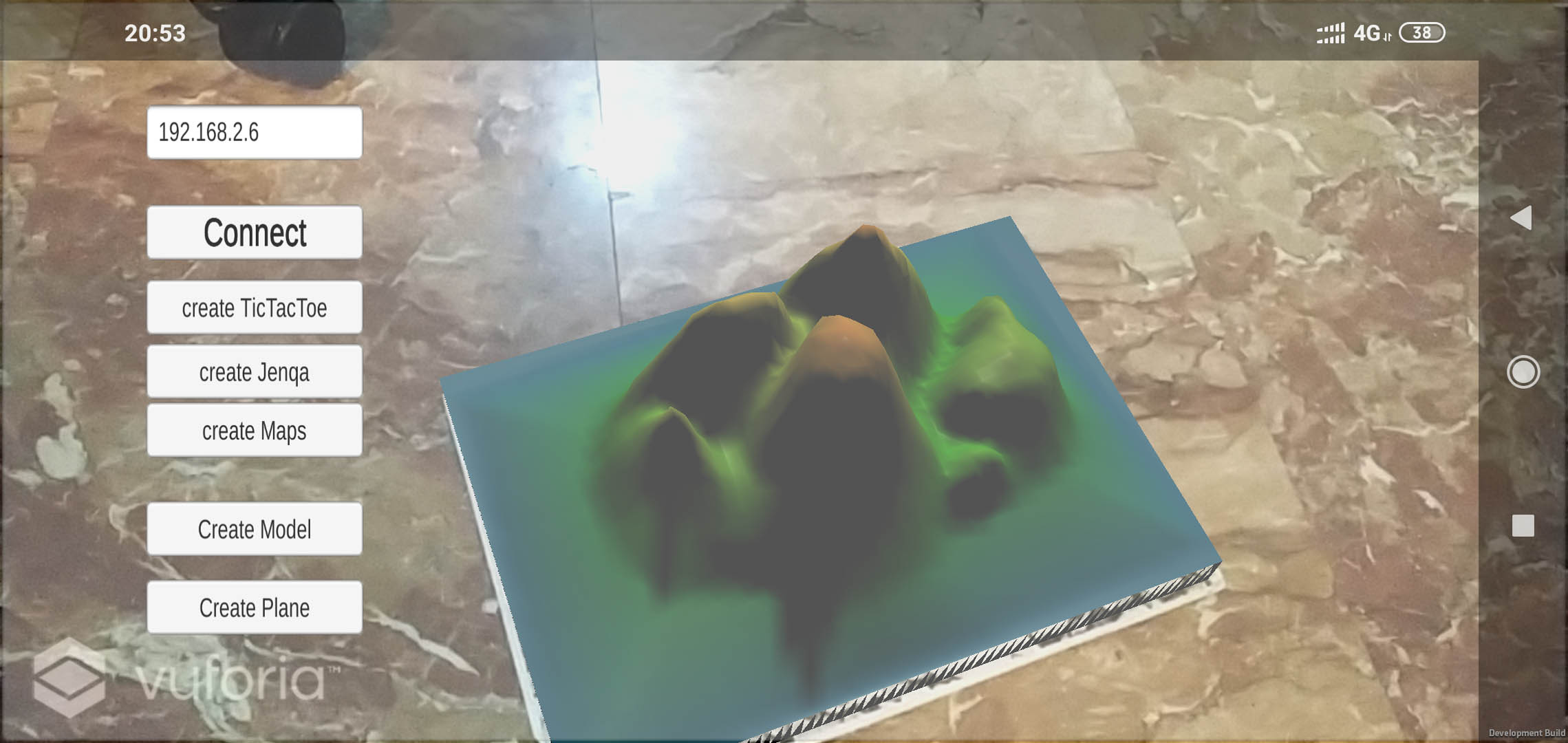}
\caption{A screenshot from the Contour Map App.}
\label{fig:MapCountour}
\end{figure}

\noindent The last game that was designed is the creation of 3D height maps (\autoref{fig:MapCountour}) using our tangible interface. In this application we are able to create mountains and valleys, building our terrain. Also in this particular application, we have the ability to create and then process 3D terrains from real height maps after an image process running at raspberry creating the 3d mesh. This process is based on the previous work of \cite{panagiotopoulos2017generation}. In this particular application, we are showing the advantages of having the 3D coordinates giving us the ability to set much more complicates commands such as setting the height.

A video clip with these applications is available at  \url{ https://www.youtube.com/watch?v=OyU4GOLoXnA} .

\section{\uppercase{Discussion \& Conclusion}}
\label{sec:conclusion}
\noindent
This work was a proof of concept that a marker-based low-cost AR with 3D TUI running in real-time (25-30 fps) is feasible to implement and use it as a testbed for identifying various problems and investigate possible solutions. If we add one or more input devices with a different color and/or different shape, then the current implementation is scalable to co-located collaborative AR, supporting two or more users. The knowledge in real-time of the 3D position of the input device offers a powerful and flexible environment for interactivity with the digital world of AR. 

Some advantages of the system are Fast 3D Registration Process, Fast Corner Detection Algorithm,  Depth Adaptive Camera Calibration, Data Fusion from RGB and Depth Camera, Simple and Fast Image segmentation,  Real-Time Operation, Versatility, 
Open Source Implementation and Hardware Compatibility.

\subsection {Future Directions}
\noindent Today, there is a lot of effort towards developing high-quality AR systems with tangible user interfaces. Microsoft-Hololense \footnote{\href{https://www.microsoft.com/en-us/hololens}{https://www.microsoft.com/en-us/hololens}} and Holo-Stylus \footnote{\href{https://www.holo-stylus.com/}{https://https://www.holo-stylus.com}} are just two of them. However, all of them are build on specialized hardware and proprietary software and there are expensive. On the other side, smartphones are continuously evolving, adding more computer power, more sensors, and high-quality display. Multi cameras and depth sensors are some of their recent additions. Therefore, we expect that it will be possible to implement all the functionalities of an AR system just in a \textit{smartphone}. In this case, computing power will be in demand. We will need to develop new fast and efficient algorithms. One way to achieve this is to make them task-specific. cMinMax is such an example, where we can find the corners of a marker (convex quadrangle) almost ten times faster than the commonly used Harris Corner Detection algorithm. The fusion of data obtained from different \textit{mobile} sensors (multiple RGB cameras, Depth Camera, Ultrasound sensor, Three‑axis gyroscope, Accelerometer, Proximity sensor, e.t.c) to locate in real-time 3D objects in 3D space and register them to the virtual world is another challenging task. A simple example is presented in \autoref{subsub:MapRealVirtual}, where we combine data from an RGB and a Depth camera in order to find the 3D coordinates of a small ball (approximated with a point) in space.   

\subsection {Conclusions} 
\noindent This paper has presented the implementation of an inexpensive single-user realization of a system with a 3D tangible user interface build with off the selves components. This system is easy to implement, it runs in real-time and it is suitable to use as an experimental AR testbed where we can try new concepts and methods. We did optimize its performance either by moving computational complexity out of the main loop of operation or by using task-specific fast procedures. cMinMax, a new algorithm for finding the corners of a markers mask, is such an example, where we have sacrifice generality in order to gain speed. 

\section*{\uppercase{Acknowledgements}}

\noindent We would like to thank the members of the Visualization and Virtual Reality Group of the Department of Electrical and Computer Engineering of the University of Patras as well as the members the Multimedia Research Lab of the Xanthi's Division of the "Athena" Research and Innovation Center, for their comments and advice during the preparation of this work. 

\bibliographystyle{apalike}
{\small
\bibliography{paperChamzasMoustakas}}

\end{document}